\documentclass[conference]{IEEEtran}
\IEEEoverridecommandlockouts

\usepackage{csquotes}
\usepackage{pdfpages}
\usepackage{graphicx}
\usepackage[style=ieee]{biblatex}
\usepackage[section,numberedsection=autolabel]{glossaries}
\makeglossaries

\newglossaryentry{function}{name=function,%
    description={\mbox{}\\*A single, self-contained piece of software, that performs a certain function}}

\newglossaryentry{feature}{name=feature,
    description={\mbox{}\\*Composed of one or more functions connected together using a certain runtime environment, usually corresponds to a certain use-case}}

\newglossaryentry{runtime environment}{name=runtime environment,
    description={\mbox{}\\*Communication middleware and virtualization mechanisms}}

\newglossaryentry{ECU}{name=ECU,
    description={Electronic Control Unit. It is an electronic device in a vehicle that is responsible for a single function}}

\newglossaryentry{TDD}{name=TDD,
    description={Test-Driven Development is a software development methodology that centers on the iterative creation of unit tests prior to the implementation of functional code~\cite{ref40:Beck2002} This approach mandates that a test case specifying the desired behavior of a code unit be written before the production code itself. As development progresses, the test suite continuously executes. New code is only written if it fulfills the requirements outlined in a failing test}}

\newglossaryentry{FDD}{name=FDD,
    description={Feature-Driven Development. It is a paradigm where the software system is iteratively developed in a series of steps, starting with an abstract model of the system, followed by extraction of a set of desired features, and ending with feature implementation and integration~\cite{ref41:Palmer2001}}}

\newglossaryentry{MBSE}{name=MBSE,
    description={Model-Based Systems Engineering is a formalized methodology within systems engineering that emphasizes using models as the primary means of information exchange and system representation~\cite{ref42:Incose2023}. This contrasts with traditional document-centric approaches. MBSE centers on creating and leveraging domain-specific models or metamodels, which capture system requirements, design, analysis, and verification elements throughout the development lifecycle}}

\newglossaryentry{contract}{name=contract,
    description={\mbox{}\\*Design by contract is a software development methodology that emphasizes the explicit definition of formal contracts between software components~\cite{ref43:Mitchell2002}. These contracts specify preconditions (what must be true before a component is used), postconditions (what must be true after execution), and invariants (conditions that must always hold true). Design by contract can be enforced through runtime assertions, unit tests, or even integrated into a programming language's syntax. This approach enhances software reliability, eases debugging, and facilitates code comprehension}}

\newglossaryentry{metamodel}{name=Metamodel,
    description={Defines the language of system description by specifying abstract entities that are part of the system, a set of possible relations between them, and their attributes}}

\newglossaryentry{instance model}{name=Instance model,
    description={A model generated from the given metamodel, populated with actual objects with concrete attribute values; an implementation of the system described in the language of the metamodel}}

\newglossaryentry{OMG}{name=OMG,
    description={Object Management Group}}

\newglossaryentry{LLM}{name=LLM,
    description={Large Language Model}}

\newglossaryentry{OCL}{name=OCL,
    description={Object Constraint Language}}

\newglossaryentry{RACE}{name=RACE,
    description={Centralized Platform Computer Based Architecture for Automotive Applications}}

\newglossaryentry{Ecore}{name=Ecore,
    description={Language of the metamodel used in Eclipse Modeling Framework}}

\newglossaryentry{OEM}{name=OEM,
    description={Original Equipment Manufacturer}}

\makeatletter 
\newcommand{\linebreakand}{%
  \end{@IEEEauthorhalign}
  \hfill\mbox{}\par
  \mbox{}\hfill\begin{@IEEEauthorhalign}
}
\makeatother 

\title{Towards Single-System Illusion in Software-Defined Vehicles -- Automated, AI-Powered Workflow \\\vspace*{20pt} \normalsize  \today{}}

\author{\IEEEauthorblockN{Krzysztof Lebioda\IEEEauthorrefmark{1}}
\IEEEauthorblockA{email: krzysztof.lebioda@tum.de \\
orcid: 0000-0002-7905-8103}
\and
\IEEEauthorblockN{Viktor Vorobev\IEEEauthorrefmark{1}}
\IEEEauthorblockA{email: vorobev@in.tum.de \\
orcid: 0000-0002-0473-148X}
\and
\IEEEauthorblockN{Nenad Petrovic\IEEEauthorrefmark{1}}
\IEEEauthorblockA{email: nenad.petrovic@tum.de \\
orcid: 0000-0003-2264-7369}
\linebreakand
\and
\IEEEauthorblockN{Fengjunjie Pan\IEEEauthorrefmark{1}}
\IEEEauthorblockA{email: f.pan@tum.de \\
orcid: 0009-0005-8303-1156}
\and
\IEEEauthorblockN{Vahid Zolfaghari\IEEEauthorrefmark{1}}
\IEEEauthorblockA{email: v.zolfaghari@tum.de \\
orcid: 0009-0004-0039-6014}
\and
\IEEEauthorblockN{Alois Knoll\IEEEauthorrefmark{1}}
\IEEEauthorblockA{orcid: 0000-0003-4840-076X}
\linebreakand
\IEEEauthorrefmark{1}\IEEEauthorblockA{Technical University of Munich (TUM) \\
School of Computation, Information and Technology (CIT) \\
Chair of Robotics, Artificial Intelligence and Embedded Systems}

}

\date{\today}

\begin{document}
\null%
\includepdf{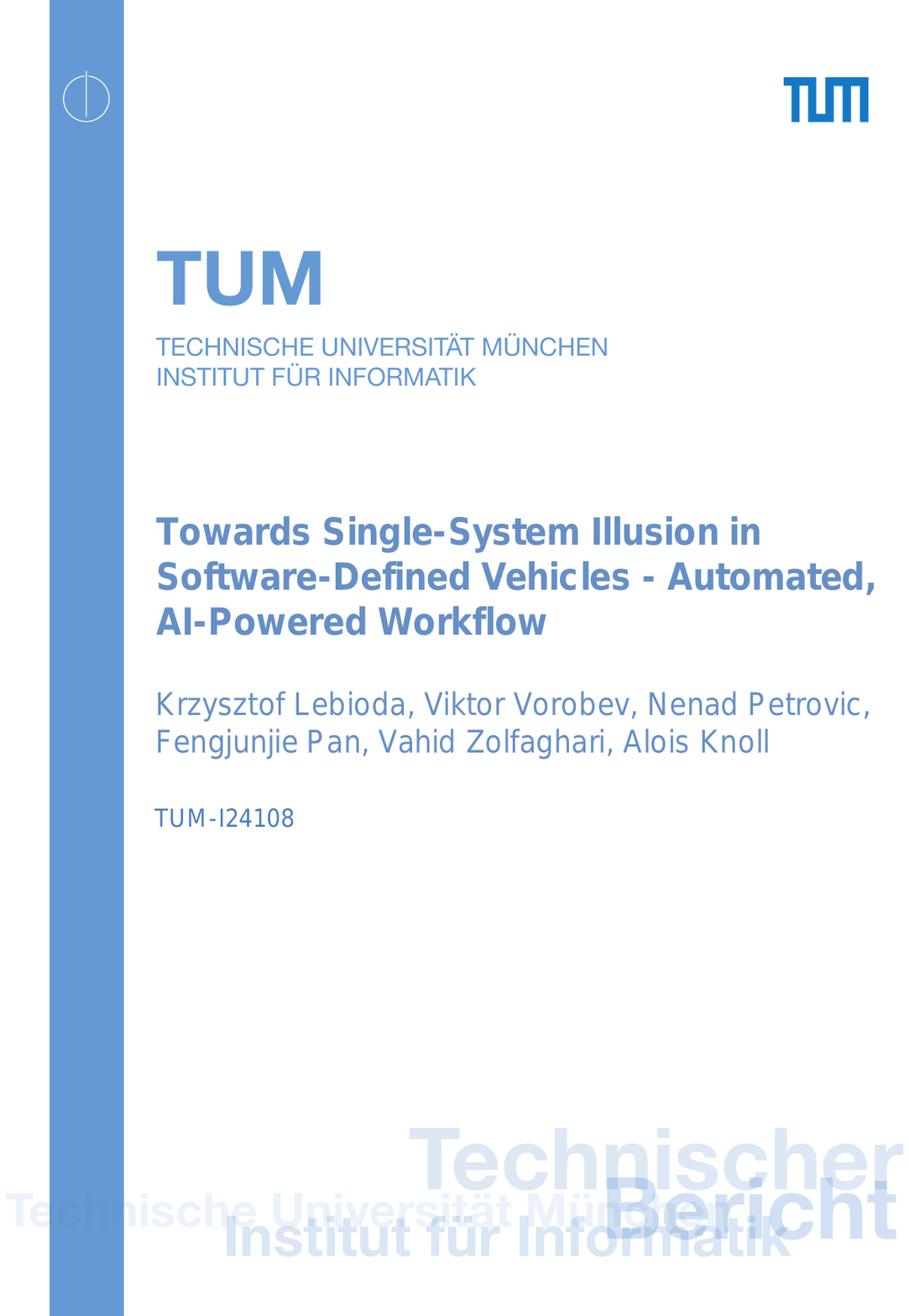}

\maketitle

\begin{abstract}
We propose a novel model- and feature-based approach to development of vehicle software systems, where the end architecture is not explicitly defined. Instead, it emerges from an iterative process of search and optimization given certain constraints, requirements and hardware architecture, while retaining the property of single-system illusion, where applications run in a logically uniform environment. One of the key points of the presented approach is the inclusion of modern generative AI, specifically Large Language Models (LLMs), in the loop. With the recent advances in the field, we expect that the LLMs will be able to assist in processing of requirements, generation of formal system models, as well as generation of software deployment specification and test code. The resulting pipeline is automated to a large extent, with feedback being generated at each step.
\end{abstract}

\section{Introduction}

The costs of vehicle software development are rising at a high rate. It is estimated that the development efforts for various software components, including OS, middleware and infotainment, but also implementation of \gls{function}s and their integration, will double in 2030 as compared to 2020~\cite{ref21:McKinsey}.

Classical software development paradigms are very rigid and slow to adapt to the rising system complexity. V-model, being the de facto industry standard, is very inflexible, lacking early feedback mechanisms, and with costly scope adjustments~\cite{ref22:Kumar2014}. Combined with highly standardized architectures like AUTOSAR~\cite{ref39:Staron2017}, this leads to very long development cycles. On the one hand, standardization and code reuse are certainly advantageous. On the other hand, over-engineering, complexity and steep learning curve are pointed out as the main problems with these well-established frameworks~\cite{ref20:Martinez2015}.

Software-defined vehicles are becoming the new trend in the automotive industry, where the functionality of the car is defined, updated and modified mainly by changes in the software. This trend affects both intra-vehicular networks~\cite{ref02:Islam2021,ref03:Correia2017,ref04:Ku2014}, as well as the internal car systems~\cite{ref05:Haeberle2020}. The increasing demand for various \gls{feature}s and software components means that the \gls{OEM}s are not able to provide the full software stack anymore. Instead, third-party software is starting to play an increasingly significant role~\cite{ref08:SBInsights2023}.

One of the trending paradigms of software development, hardware modeling and resource allocation is model-based system engineering (\gls{MBSE})~\cite{ref32:Ambrosio2017,ref33:Pohlmann2019,ref34:Azzoni2021,ref16:Pan2023}. Coupled with the principles of design by \gls{contract}, where the obligations of all interacting components are written down in a formal way, \gls{MBSE} becomes a powerful tool that enables software-defined vehicles.

With the advent of Large Language Models (LLMs), new automation possibilities are opening. We would like to leverage the generative power of modern AI to define a new software development paradigm, which goes beyond the current standards, and which will be easily extensible in the future.

The presented approach to software development draws from agile principles, like feature-driven development (\gls{FDD}), test-driven development (\gls{TDD}), and low-code, model-driven development. Combining different techniques allows addressing the complexity of the problem, developing rapidly, and providing necessary feedback that is used to improve the overall system design. This fits well with the software-defined vehicle paradigm, where the system keeps evolving by updating and modifying mainly the software components.

When using the proposed workflow, software designers and developers should perceive the system as a single logical entity -- single-system illusion. Different software modules are not bound to the underlying OS, middleware, or hardware topology. The \gls{RACE} project is an example of early designs that show how such a system may look like~\cite{ref17:Sommer2013}. This is in contrast with the current systems, where the functions are spread to many different \gls{ECU}s, and are oftentimes implemented in a way that accounts for the inter-component connections, which breaks modularity. 

\begin{figure*}[t]
    \centering
    \includegraphics[width=\textwidth]{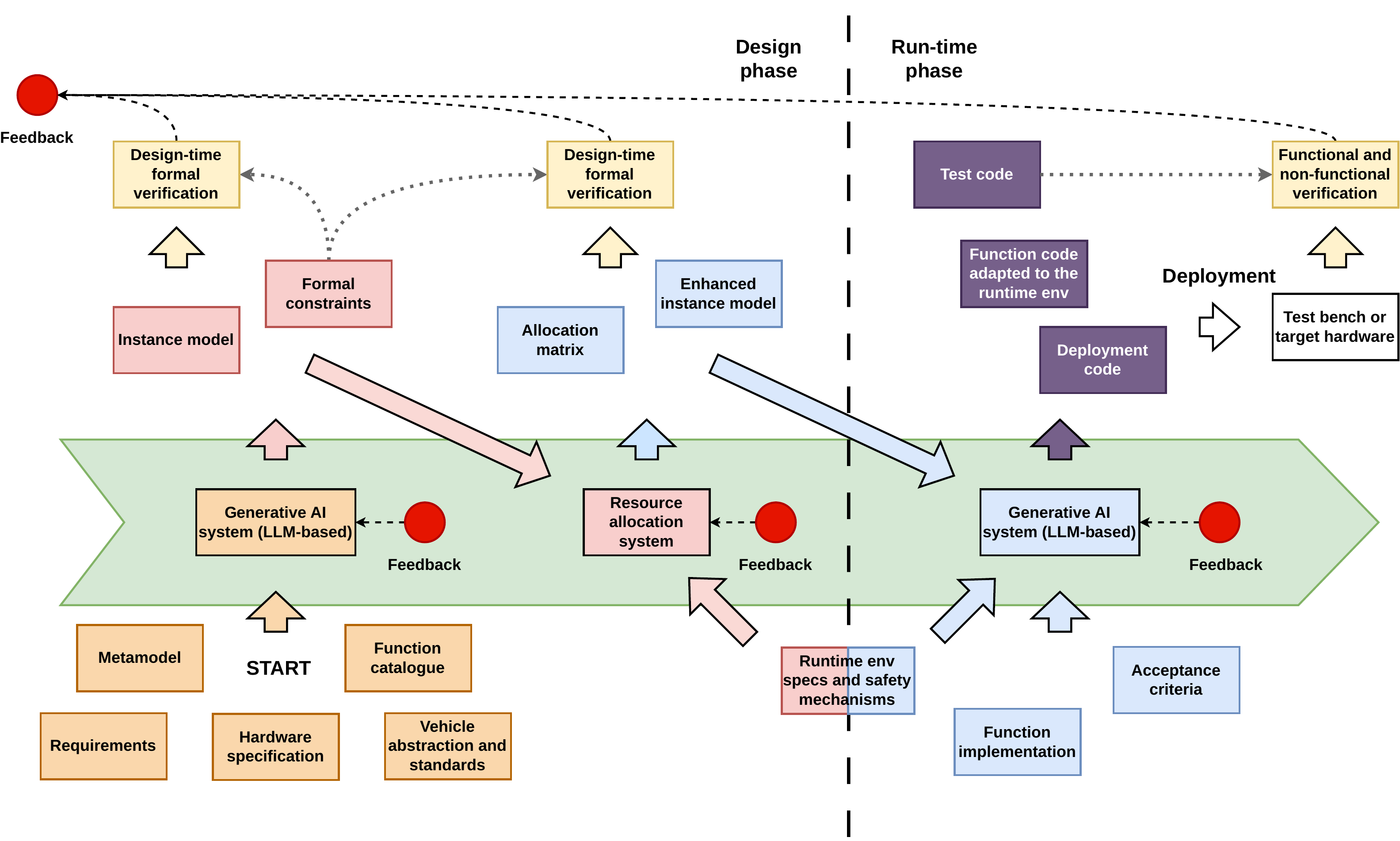}
    \caption{Proposed workflow with generative AI in the loop. Steps are color-coded: 1) Inputs in orange, outputs in red - generation of the instance model and formal constraints. The formal constraints are used for automatic verification of the instance models; 2) Inputs in red, outputs in blue - resource allocation, which produces an allocation matrix and enhances the instance model with details about software-to-hardware mapping; 3) Inputs in blue, outputs in purple - code generation. The tests generated during this step are used for automatic testing of the deployed system; 4) Yellow - validation and verification, which is performed as part of the other steps, and is not a separate step as such.}
    \label{fig1}
\end{figure*}

In order to achieve the goal of providing a single-system illusion, the various components of the software stack must be clearly separated. Applications, middleware, OS and hardware must be sufficiently modular to allow modifications of one layer without the need to totally rewrite the others. Again, \gls{RACE} has shown how to clearly delineate applications from the \gls{runtime environment}~\cite{ref17:Sommer2013,ref18:Becker2015} and even how to hide specific safety mechanisms from the service-level developers~\cite{ref18:Becker2015}. However, the resulting system was strictly bound to the runtime environment with a single message exchange paradigm (publish-subscribe). We would like to go beyond and allow the designers to choose their middleware without such constraints, possibly even allow for multiple middlewares within the same system. We also strive for a system where certain general safety mechanisms, like process redundancy, watchdogs and monitoring, or data integrity checks, are separated from the applications and the runtime environment and where they can be applied to other components as required.

The innovative aspect of the workflow which we propose, is the use of modern AI, specifically \gls{LLM}s, in the development process. The goal is to use AI in synergy with model-driven development approaches, where formal system information, including the system model (so-called instance model) and constraints in formal languages, is generated from the available components' (software and hardware) description, and from the set of requirements, both functional (which functions to include) and non-functional (safety, performance). Based on the safety requirements, functions have different safety measures (redundancy, failure handling mechanisms) applied to them. Formal verification is performed on the instance model relying on a rule-based approach, such as Object Constraint Language (OCL)~\cite{ref30}. This way, it is possible to prove that logical connections between elements are correct while the properties of distinct elements are within the desired bounds.

The generated model is used as a base for an automatic resource allocation method, which takes into consideration the hardware and software specifications as well as allocation constraints. Flexibility of the optimization algorithm must be emphasised, as it ought to support both test environments and the target vehicle architectures. It must also be able to generate an optimal configuration in the Pareto sense, i.e., based on the chosen optimization criteria, like maximal performance or minimal power consumption.

The resulting allocation matrix together with the instance model become inputs to a code-generating system based on generative AI. The functional model is combined with information about the desired runtime environment, acceptance criteria, and the list of available software functions from the software catalog. The system outputs working code that can be deployed to the desired architecture. The process of code generation includes parametrization of deployment code, e.g., with connection details like addresses and ports, generation of adapter code that translates data from function-specific formats into middleware primitives, but also tests that will be used for functional and non-functional verification. The synthesized architecture can be verified in simulation against the non-functional requirements, e.g., by using fault injection into the stream, similar to \gls{RACE}~\cite{ref18:Becker2015}.

\section{Methodology}

The workflow was designed with the following principles in mind: short development cycles, automated feedback at each step, focus on modularity and flexibility, and automated generation of boilerplate code. A number of artifacts is produced as a result of each step. All of these artifacts should be available to the users in a human-readable form. However, they should also have a machine-readable format, which can be used for automatic verification. Verification at each step provides early feedback, and should keep the development cycles short. The software components at all levels of abstraction should be modular to allow the AI tools to easily (re)generate the boilerplate code as needed.

The proposed workflow is presented on Fig.~\ref{fig1}. It consists of two phases -- the design phase, and the run-time phase. During the design phase, formal descriptions of the desired system in the form of models and constraints are generated, based on the provided requirements, standards, available software and hardware components etc. During the run-time phase, glue code, tests and deployment descriptors are generated from the model. The software is deployed and tested either on the test bench, or on the target architecture.

\subsection{Feature-driven and model-driven development}
The developers should first and foremost focus on the features that are supposed to be deployed to the target vehicle hardware. At a very abstract level it can be achieved by writing a set of requirements in natural language. A generative AI in the form of \gls{LLM} is used to process these abstract ideas and put them against the available software functions, hardware specifications, safety standards and vehicle abstraction specifications. The \gls{LLM} outputs an instance model, which is based on the given metamodel and which abstracts the target system, and formal constraints that are derived from requirements. The generated instance model represents how the function graph is connected, what the requirements are for all nodes, as well as what properties the target hardware has. At this point, the designers should not be concerned about the details of virtualization, operating systems, or the precise mapping between software and hardware.

\noindent
\\
\textbf{Inputs:}
\begin{itemize}
    \item \emph{Metamodel} -- an abstract language of system description. Metamodel does not describe any particular system. Instead, it is a template that must be populated, creating a so-called instance model. One way the meta- and instance models can be written is the Object Management Group (\gls{OMG}) standard. When it comes to implementation, we rely on \gls{Ecore}~\cite{ref31:EMF2009} within Eclipse Modelling Framework (EMF), which provides the tools for metamodel design, model instance creation and API for model instance manipulation. Other modeling language, like SpesML, may be used, depending on the desired toolchain.
    \item \emph{Requirements} -- a list of requirements for the system in free-form text. Requirements contain basic system description, constraints, etc. Although the process of gathering, extraction and improving of the requirements set is an interesting research problem on its own, we are not going to address it in this document. However, we do intend to assess the quality of models generated by the \gls{LLM}s using existing rule-based systems, and test cases where requirements are contradictory or incomplete.
    \item \emph{Vehicle abstraction and standards} -- essential for defining vehicular systems, for example, ISO 26262~\cite{ref35:ISO26262}, ISO 23150~\cite{ref36:ISO23150}, Vehicle Signal Specification~\cite{ref37:VSS}.
    \item \emph{Function catalogue} -- a list of available software components (functions), interface specifications, and requirements regarding computational power, power consumption, etc.
    \item \emph{Hardware specification} -- specifications of the available hardware. The specification could be in the form of a graph, data sheet, or potentially another instance model.
\end{itemize}

\noindent
\textbf{Outputs:}
\begin{itemize}

    \item \emph{Instance model} -- an instance of the metamodel, which describes the connections between all involved functions. The function nodes are annotated with their requirements in terms of safety, required power, etc. At this point in the design phase, the hardware graph is included in the model, but the details of mapping of software to the hardware are missing. The model will be enhanced with full mapping in the following step -- resource allocation. As in the case of the metamodel, we make use of \gls{Ecore} for the implementation, but other modeling languages may also be used.
    \item \emph{Formal constraints} -- a set of logical constraints that must be satisfied within a user-provided model instance. These rules are typically derived from a reference architecture, requirements or relevant ISO standards. In our approach, we employ Object Constraint Language (OCL) to express and enforce these verification rules.
\end{itemize}

\noindent
\textbf{Verification and feedback:}
\begin{itemize}
    \item Feedback includes testing that interfaces of different components are compatible and meet the criteria specified in the formal \gls{contract}. Failure to adhere to this contract may indicate that the chosen components are incompatible, or that the interface specification is incorrect.
    \item Violations of formal constraints within the scope of user-created specification (including component quantity, property values, or inter-component relationships) will result in model verification failure. The system will provide the user with natural language suggestions outlining necessary modifications. To generate these suggestions, the \gls{LLM} will translate the verification results into actionable guidance.
\end{itemize}

\subsection{Resource allocation}
During this step, the generated instance model and formal constraints are translated into a format recognizable by the chosen mapping and optimization algorithm. The algorithm should be flexible to support various optimization criteria to develop a Pareto-optimal solution based on the selected goals (cost, performance, power use, etc.). Existing approaches to mapping and optimization offer many possibilities: from classical optimization techniques like integer linear programming~\cite{ref01:Askaripoor2021,ref44:Pan2022} through genetic algorithms~\cite{ref10:Kao2020,ref11:Weng2008}, to the use of graph neural networks~\cite{ref13:Schuetz2022,ref14:Cappart2021, ref15:Rusek2020}.

\noindent
\\
\textbf{Inputs:}
\begin{itemize}
    \item \emph{Instance model} -- model generated in the previous step.
    \item \emph{Formal constraints} -- formal constraints generated in the previous step. They are derived from requirements and contain restrictions and rules for resource allocation.
    \item \emph{Runtime environment specification} -- specification of the chosen communication middleware (ROS, ZeroMQ), available message passing paradigms, and virtualization mechanisms (Docker, VM).
    \item \emph{Available safety mechanisms} -- abstract description of safety mechanisms, like redundancy, failure masking by voting, etc. These will later be instantiated according to the chosen runtime environment and available software components (\gls{function}s).

\end{itemize}

\noindent
\textbf{Output:}
\begin{itemize}
    \item \emph{Allocation matrix} -- mapping between software and hardware components.
    \item \emph{Enhanced instance model} -- instance model generated in the previous steps, but enhanced with detailed software-hardware mapping.
\end{itemize}

\noindent
\textbf{Validation and feedback:}
\begin{itemize}
    \item Similar to the previous step, validation includes testing of the enhanced model against a set of \gls{OCL} rules covering redundancy aspects and safety-critical constraints.
\end{itemize}

\subsection{Code generation and deployment}
In this step, another instance of generative AI is used to put all available pieces together into working code. Techniques like discriminative reranking~\cite{ref25:Lee2021,ref26:Li2024}, application of verifier models~\cite{ref27:Ni2023,ref28:Li2023} or self-collaboration~\cite{ref29:Dong2023} may be applied to improve the code quality. The generated code may include (but is not limited to) glue code for various functions, like adapters from function interfaces and middleware interfaces, deployment files parametrized with proper addresses, ports and including process redundancy, injection of test code into the pipelines, and actual test cases~\cite{ref23:Chen2022,ref24:Liu2023} based on acceptance requirements.

\noindent
\\
\textbf{Input:}
\begin{itemize}
    \item \emph{Allocation matrix} -- mapping of software to hardware generated in the previous step
    \item \emph{Runtime environment specification} -- specification of the chosen communication middleware (ROS, ZeroMQ), available message passing paradigms, and virtualization mechanisms (Docker, VM).
    \item \emph{Acceptance criteria} -- functional and non-functional criteria used for the generation of test code and test cases.
\end{itemize}

\noindent
\textbf{Output:}
\begin{itemize}
    \item \emph{Deployment files} -- adapted to the particular architecture and runtime environment
    \item \emph{Test cases} -- automatically generated test cases that can be used to validate both functional and non-functional requirements.
    \item \emph{Function code} with proper wrapping (Docker) and adapter code for data translation.
\end{itemize}

\noindent
\textbf{Verification and feedback:}
\begin{itemize}
    \item Execution of the acceptance tests, either in a simulated environment, or on the target architecture.
    \item Feedback is divided into functional and non-functional. Failures detected during execution of the functional test could indicate bugs in the software modules or in the integration code. Failures of the non-functional tests may indicate that certain requirements were incorrect or missing, or that the optimization criteria during the allocation phase must be modified.
\end{itemize}

\section{Scope, complementary work, limitations}
In this document we are focusing on automatic generation of software systems, integration and deployment code, and mapping to hardware. However, there are many places where complementary research can be conducted. The first issue is completeness and consistency of the user requirements. Although the instance model and formal constraints produces by the LLM in the first step are verifiable, we do not know how well the LLM will be able to handle conflicting and incomplete requirements, and how it will impact the quality of the generated artifacts. Another possible venue of research is the hardware specification and representation. In the current workflow it is assumed that this description is provided to the LLM in the form of a graph (or a semi-formal textual description), but it ought to be possible to generate the hardware model automatically from the requirements, similar to how we generate the software model. An example of Ecore model instances creation relying on ChatGPT in presented in ~\cite{ref45:Petrovic2023}.

Code generated by \gls{LLM}s is far from perfect~\cite{ref38:Spiess2024}. We are planning on focusing on certain simple use cases and building a proof-of-concept pipeline. Human supervision will definitely be needed, especially to examine correctness of the automatically generated verification code, like test cases. We expect that in the near future, the quality of code generated by \gls{LLM}s will improve massively, which will allow us to generate more complex systems, and to progressively remove the need for human supervision.

\section{Conclusion}

We propose a software development process that extends beyond the current industrial standards. In this process, the generative AI should become an integral part and progressively take care of many menial tasks. The advantage of using AI over classical tools for automatic translation is the generative power of the models, as well as the ability to understand natural language. Although the currently available \gls{LLM}s have many shortcomings, the progress in the field is blazingly fast, and their rapid evolution suggests the potential to revolutionize code generation in the coming years. Our proposed workflow is designed for extensibility and adaptability, ensuring it remains relevant amidst ongoing advancements in AI capabilities, allowing an increasing number of stages of the process to become automated.

\printglossaries

\printbibliography

\end{document}